# Noise in Nonohmic Regimes of Disordered Systems

K. K. Bardhan*, C. D. Mukherjee* and U. N. Nandi†

*Saha Institute of Nuclear Physics, 1/AF Bidhannagar, Kolkata 700 064, India
†Scottish Church College, 1 & 3 Urquhart Square, Kolkata 700 006, India

**Abstract.** We present here a short review of mainly experimental properties of noise as disordered systems are driven into non-ohmic regimes by applying voltages of few volts only. It is found that the noise does not simply follow the resistance in that the direction of change of noise could be opposite to that of resistance. It is discussed how this and other properties make the noise a complementary and incisive tool for studying complex systems, particularly its dynamic properties. Study of noise in non-ohmic regimes in physical systems is rather in a nascent stage. Some of the open issues are highlighted.

## 1. INTRODUCTION

The ubiquitous low frequency resistance fluctuations in conductors have proved in recent times to be an unique and increasingly useful tool for probing various condensed matter systems[1, 2]. For this purpose, fluctuations have been often studied also as a function of other relevent variables such as temperature. But, studies of fluctuations as a function of bias are rather scarce. In the early stage of noise study, the focus was on its fundamental properties and simple systems such as pure metals or semiconductors[3] were used. These systems remain ohmic except at very high electric fields and hence, were not particularly suitable for probing non-ohmic regimes. However, with advent of semiconducting devices that contain abundantly non-ohmic junctions such as metal-semiconductor and semiconductor-semiconductor ones, noise-related performance issues assumed much importance. Study of the noise in these non-ohmic junctions in late seventies (see Hooge et al.[4] for references) constitutes one of the earliest example of investigation of noise in non-ohmic regime. About the same time, noise measurements in a charge-density system of $NbSe_3$ as a function of electric field [5] revealed intricacies of the pinning dynamics, thereby proving its usefulness for the first time in a non-device physical system. In recent times, there are signs of increased level of activities mainly upon realisation that such studies would be uniquely useful especially in complex systems such as composites[6], mangenites[7] and others[8, 9] where usual transport measurements of average quantities prove to be inadequate. These systems have varying degree of disorder and are easily driven into non-ohmic states by application of bias of only few volts (thus, fluctuations in hot-electrons[10] remain beyond the scope of this review). Noise, being proportional to the fourth moment of the current distribution, is more sensitive to the microstructures (i.e. disorder) or onset of non-ohmic behaviour (i.e. change in conduction mechanism) than the resistance which is proportional only to the second moment of the current distribution. Recently, nonlinearity has been used by Vilar and Rubi[11] to suggest an interesting possibility of suppressing intrinsic noise in a nonlinear system by adding external noise.

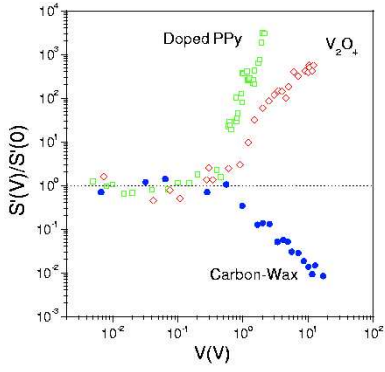

FIGURE 1: Suitably normalised relative noise as a function of bias in three systems with different conduction mechanisms. Resistances in all three cases decrease with bias. Yet, the relative noise behaves differently by decreasing in composites (carbon-wax), and increasing in other two cases. (Figure from Ref. [12]).

Let us note that Hooge's empirical formula[13] can be generalised in the following manner using chordal resistance $R = V/I$:

$$\mathscr{S} = \frac{S_R}{R^2} = \frac{S_V}{V^2} = \frac{S_I}{I^2} = \frac{V^\gamma}{f^\alpha}\mathscr{R}(R) \tag{1}$$

where $S_X = <\delta X^2>$ is the spectral density when $X = R, V, I$ is the fluctuating variable. $I$ is the current through sample when $V$ is the applied dc voltage. The value of the exponent $\gamma$ is normally zero but can assume a non-zero value in disordered systems (see next section). The function $\mathscr{R}$ depends upon the particular system under consideration. In homogeneous samples, $\mathscr{S} \sim \mathscr{R} \sim R$. The relative noise power, $\mathscr{S}$ is analogous to conductance or resistance. Naturally, one would like to know how $\mathscr{S}$ behaves vis-a-vis $R$ as a function of bias. For example, does it increase with $R$ as it does *always* in ohmic states? As seen in Fig. 1, the relative noise does not necessarily follow the resistance in the nonohmic regime. The figure shows bias-dependent relative noise in three different systems with different conduction mechanisms. Resistances of all the samples *decrease* with bias. Yet, the relative noise behaves differently, decreasing in one case and increasing in others, obviously depending upon the underlying conduction dynamics. This amply illustrates the complementary role that the noise studies can play in unravelling complex systems. Interestingly, such behaviour of noise can be seen even from very general consideration of the bias-dependent resistance

$$R(I) = R_o f(I, R_o, c_1, c_2, ...) \tag{2}$$

where $R_o$ is the resistance at zero bias, $f$ is a function such that $f(0,..) = 1$. $c_i$'s are various relevant parameters characterising the non-ohmic regime. Its fluctuation $\mathscr{S} = <\delta R^2>/R^2$ is then given by

$$\mathscr{S} = s_1 + \sum_i \left(\frac{\partial \ln f}{\partial c_i}\right)^2 <\delta c_i^2> + \sum_{i \neq j} \frac{\partial \ln f}{\partial c_i}\frac{\partial \ln f}{\partial c_j} <\delta c_i \delta c_j> + \sum_i \frac{\partial \ln f}{\partial c_i}\frac{\partial \ln f}{\partial R_o} <\delta c_i \delta R_o> \tag{3}$$

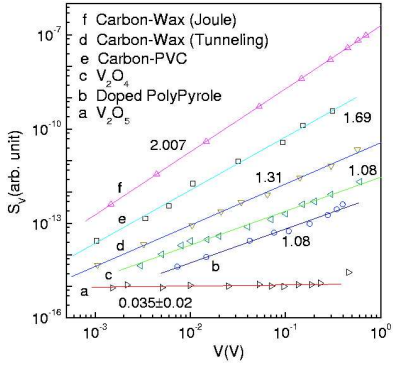

FIGURE 2: Noise voltage power vs. voltage in ohmic regimes of several disordered systems. The bias exponent $\beta$ in each system is given by the slope of the corresponding curve as indicated. Notice that $\beta$ values range from almost zero ($V_2O_5$, curve *a*) to 2 (carbon-wax composite, curve *f*). (Data of curves *b*,*c*,*d* from Ref. [12] and those of *f* from Ref. [15]).

where $s_1 = \mathscr{S}_o(1+(\partial \ln f/\partial \ln R_o)^2)$ and $\mathscr{S}_o = <\delta R_o^2>/R_o^2$ is the relative noise in the ohmic state. If there is no correlation among $c_i$'s and $R_o$ it follows from above that the noise in non-ohmic regimes is *always* greater than that in ohmic state irrespective of whether $f$ increases or decreases with bias. On the other hand, if any correlation does exist the noise may increase or decrease depending on the signs of the correlation terms on right hand side of Eq. (3). In practice, it may be a challenge to identify the appropriate set of parameters $c_i$'s in physical systems. If $f$ is known the corresponding noise could be calculated using Eq. (3). Such a method, for example, was used by Carbone et al.[14] to calculate the photocurrent noise.

We discuss below behaviour of noise in nonlinear regimes of various disordered systems and how the latter provides physical information. However, we first describe an anamolous behaviour in bias exponent ($\gamma \neq 0$ in Eq. (1) ) that seems to be observed *only* in the ohmic regimes of disordered systems.

## 2. OHMIC REGIME REVISITED

In normal conductors like metals or semiconductors, numerous experiments [13] have verified that the bias exponent $\beta(=2+\gamma = \partial \ln S_V/\partial \ln V)$ is equal to 2. In the ohmic state, it is an identity resulting from the assumption that resistance fluctuations exist *independent* of any current but become 'visible' as voltage fluctuations when a constant current $I$ is passed: $<\delta V^2> = I^2 <\delta R^2> \sim V^2$. But in case of disordered systems, $\beta(\gamma)$ even in the ohmic states are often found to differ from the value of 2(0) in the ordered systems as seen in the Fig. 2. The systems in the figure range from polaronic ($V_2O_5$) to Mott (Doped Polypyrol,$V_2O_4$) to composites (carbon-wax, carbon-PVC). Carbon-wax samples in the figure differ only in carbon fraction. One with $\beta=2$ is much less disordered than the one with $\beta = 1.3$. In fact in composites, there is a clear trend of $\beta$ decreasing with increasing disorder (or decreasing conducting fraction) [12, 16]. The lower bound for $\beta(\gamma)$ is obviously 0(-2):

$$0 \leq \beta \leq ? \tag{4}$$

There is no theoretical upper bound of $\beta$. A value as high as 4 has been reported in discontinuous Pt films[17]. As per available data, there seems to be a preponderance of $\beta$ less than or equal to 2 (i.e. $\gamma \leq 0$)[18, 12, 16, 15]. However, there are several systems[19, 17, 20] where $\gamma$ varies from negetive to positive values depending on sample structures. Till now, a proper understanding of such anamolous values of $\beta$ is lacking. Interestingly, similar anamolous dependence of shot noise on bias has also been predicted in mesoscopic systems[21] in that $S/2eI$ is no longer a constant but a function of bias. Here, the electron-electron interaction has been shown not to disturb the ohmicity but affect the current fluctuations.

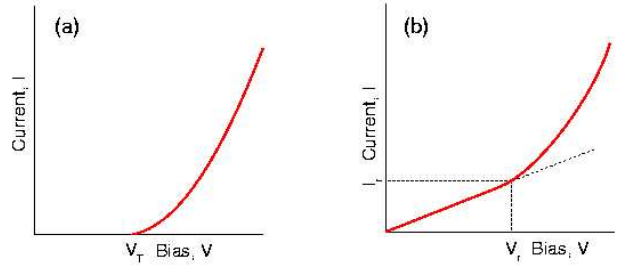

**FIGURE 3.** Schematic I-V curves corresponding to two possible modes of onset of nonlinearity: one (a) with a threshold, and another (b) without.

To take into account generally non-zero $\gamma$ in disordered systems it is useful to generalise the definition of relative noise as used in Fig. 1: $\mathscr{S}' = \mathscr{S}/\mathscr{V}^\gamma$. Some authors (e.g. C. Parman et al.[20]) termed the noise as nonlinear in case of $\gamma \neq 0$. However, the term 'nonlinear noise' is used here to describe, in analogy with nonlinear conductance, bias-dependent noise.

## 3. ONSET OF NONLINEARITY

One of the important characteristics of nonlinearity is the mode of its onset. Accordingly, systems could be divided broadly into two groups (Fig. 3). One group of systems requires application of a finite bias $V_T$ before any current can flow through samples (Fig. 3a) whereas in others, there is no apparent threshold for conduction (Fig. 3b). In the latter, one can still define a bias scale $V_r$ (or alternatively, a current scale $I_r$) for onset of nonlinearity or simply, a scale of nonlinearity. Noise invariably increases rapidly with the onset of current in systems with finite thresholds. But, in systems without thresholds, the relative noise can either increase or decrease with bias as illustrated in Fig. 1.

Another aspect of the onset of nonlinearity involves the question whether noise and resistance start deviating from their respective ohmic behaviour at the *same* value of bias or not. Let $V_n$ and $V_r$ denote suitably defined values of bias for onset of nonlinearity for noise and resistance respectively in a system. Considering the fact that noise is more sensitive to disorder than resistance one would expect that $V_n$ will be always less than or equal to $V_r$. Thus, the onset ratio $b = V_r/V_n$ is given by

$$b \geq 1 \qquad (5)$$

This is seen in Fig. 4 which presents experimental $I-V$ and $S_V-V$ characteristics of a space-charge-limited-current(SCLC) diode.

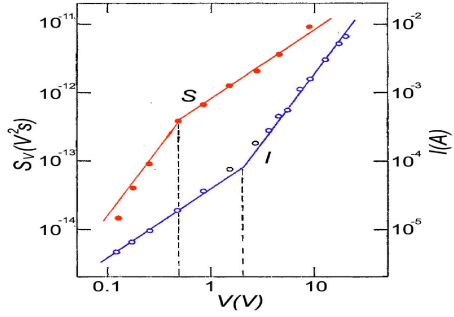

FIGURE 4: Experimental evidence that the onset ratio $b = V_r/V_n$ (about 5) is greater than 1. Curves are $I - V$ and $S_V - V$ characteristics of a SCLC diode. Figure from Ref. [4].

## 4. NOISE CHARACTERISTICS: $\mathscr{S}' - V/I/R$

While presentation of noise power as a function of either current or bias is straightforward, that as a function of (current-dependent) resistance requires furthur clarification as $\mathscr{S}' - R$ is not unique. Noise depends strongly on the methods used for changing the resistance. This is seen in Fig. 5 where the composite samples have their resistances varied in two different ways - by varying static disorder (i.e. fraction of conducting component) (a) and by Joule heating (b). The change in (non-ohmic) resistance of a sample in case of Joule heating is much less than the change in ohmic resistance caused by varying disorder but is accompanied by a huge increase in the noise compared to the other case.

Figs. 1 (composite) and 4 (SCLC diode) provide two different examples of monotonically varying noise characteristics in absence of any phase change. However, if there is any field-induced phase change the noise instead usually goes through a peak as in Fig. 7. Phases being referred to are not the usual thermodynamic ones, but rather ones characterised by different transport properties.

### 4.1. Normal systems (without any phase transition)

A typical characteristics posesses an initial linear part, then departs from linearity at some voltage $V_n$. Like nonlinear conductivity, the nature of the noise characteristics is determined by the physics of the given system. To our knowledge, composites are the most extensively studied system in non-ohmic regime[12, 15]. Limited experimental data that are presently available tend to suggest that there could be general relations between the generalised relative noise power $\mathscr{S}'$ and the bias-dependent resistance $R$. When both are considered togeather one comes across four different scenerios according to the relative directions of changes as given in Table I.

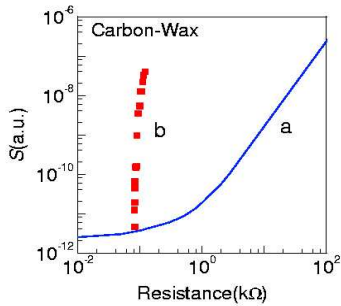

FIGURE 5: Comparision of noise in a composite system as a function of resistance when the latter is varied in two different ways. Curve *a* is obtained from samples with different degree of static disorder (ohmic resistance) while curve *b* was obtained from a sample subjected to Joule heating (non-ohmic resistance). (Figure from Ref. [15]).

**TABLE 1.** Four types of relative behaviour of resistance and noise in non-ohmic regimes of various systems. $\Delta R$ and $\Delta \mathscr{S}'$ represent changes from ohmic values of resistance and generalised relative noise respectively. $\mathscr{R}$ is the function as defined in Eq. (1). ↑ and ↓ indicate increase and decrease respectively. See text for other symbols.

| $\Delta R$ | $\Delta \mathscr{S}'$ | System | $\mathscr{R}$ | Ref. |
|---|---|---|---|---|
| ↓ | ↓ | Composites ($p \geq p_c$) | $R^{w_V}$ | [12] |
|   |   | Planar lipid membrane |   | [22] |
|   |   | SCLC diode |   | [4] |
|   |   | 1D Charge-transfer salts | ? | [23] |
| ↓ | ↑ | Variable Range Hopping | $\exp(R_o/R)^{w_V}$ | [12] |
|   |   | Solid lipid layers |   | [24] |
|   |   | ZnO (varistar) | ? | [25] |
| ↑ | ↑ | Composites ($p \gg p_c$), | Polynomial | [15] |
| ↑ | ↓ | ? | ? |   |

*R decreasing, $\mathscr{S}'$ decreasing.* The first three systems exhibiting this type of variations are quite diverse in nature, yet possess the same power-law dependence (see Fig. 6):

$$\mathscr{S}' \sim R^{w_V} \quad (6)$$

Here $w_V$ is an exponent characteristic of the system in question. It is 1 in SCLC diode[4], about 3 in carbon-wax composites (low conducting fraction, $p \geq p_c$, $p_c$ being the percolation threshold) [12] and 1.5 in planar lipid membrane. The value is well explained within the theoretical modelling of SCLC diode[26]. But there is no theoretical understanding for other systems. A question naturally arises here regarding any general feature that is common to all the three systems, and that could explain the reduction of noise? It turns out that in all these systems, increasing bias leads to progressive increase in number of fluctuators (injected charges in diode, opening of new conducting channels in composites or pores in membranes). The increase in number of fluctuators, in turn, leads to decrease in noise. However, this does not explain the power-law in (6).

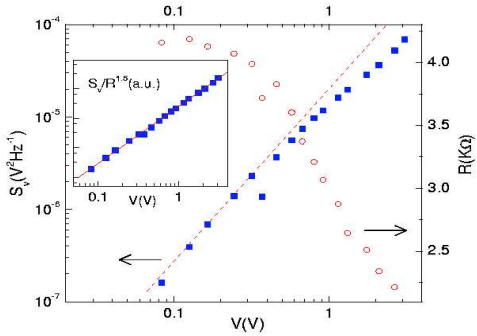

FIGURE 6: Illustration of power-law function in planar lipid membrane. The main panel shows voltage noise power and resistance as a function of voltage. $S_V$ deviates from the initial $V^2$ variation in the non-ohmic regime but $S_V/R^{1.5}$ becomes a straight line proportional to $V^2$ (inset). (Figure adapted from Ref. [22]).

*R decreasing, $\mathscr{S}'$ increasing.* This type of variation is observed in Mott systems (see Fig. 1) and others. Comparision with composites is particularly striking since the percolative picture is invoked in both type of systems. An emperical $\mathscr{R}$ is given by

$$\mathscr{S}' \sim \exp(R_o/R)^{w_V} \tag{7}$$

*R increasing, $\mathscr{S}'$ increasing.* The Joule regime in composites with high $p \gg p_c$[15] is an example of this type of systems. Nonlinearity in both resistance and noise here is well understood. The noise in this case is a polynomial of $R$ in contrast to having a power-law dependence in composites near $p_c$. Recently, Pennetta and her coworkers[27] have carried out extensive simulation of noise in the Joule regime of a random resistor network under two competing processes, driven by bias and thermally activated.

*R increasing, $\mathscr{S}'$ decreasing.* No system exhibiting this type of variation is known.

### 4.2. Phase Transitions: Noise peaks

*Systems with threshold.* A topic that has seen much activities in recent times is one of collective motion in disordered systems. Such a collective motion is usually characterised by a finite threshold value of the driving force. Examples of systems exhibiting collective motions include charge density waves (CDW's) [8], flux line lattice in type II superconductors[9], Wigner crystals in semiconductors. A finite threshold is essentially due to presence of pinnings in the system that inhibit the motion until the applied bias is sufficient to overcome the pinning force. As expected, noise measurements in these systems[28, 30, 31] have either confirmed earlier conclusions or yielded new information on dynamic quantities such as coherence lengths of moving entities. For example, the narrow-band noise in CDW's reinforces the picture of sliding CDW's by touching on the details of its periodic interaction with pinning potential. Fig. 7a shows a typical broad-band noise curve in a CDW system. Noise increases rapidly beyond the threshold field and then goes through a maximum or peak. The authors[28] ascribed the noise to fluctuations between metastable CDW states with random pinning forces. The peak in the figure appears to represent a rather generic feature. It may indicate a transition from the initial noisy state due to jerky motion at the onset to a quieter phase. Indeed, at high field the transport should be less sensitive to the pinnings and consequently, less noisy. In type II superconductors, such structures[30] have been interpreted as a field-induced transition from a quiet elastic flow state of the flux flow lattice to a noisy plastic flow regime corresponding to peak region and finally, to a (molten) fluid flow regime again with low noise.

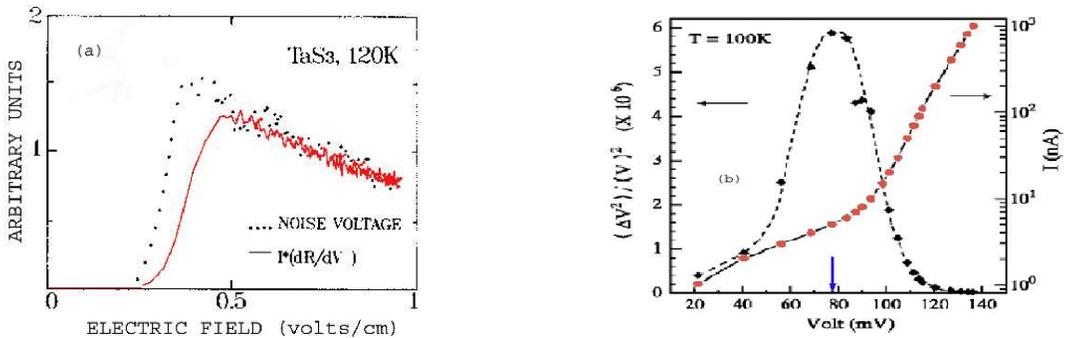

**FIGURE 7.** a) Noise in charge density wave system of TaSe$_3$ as a function of bias. The increase of noise coincides with appearance of current flow in the sample. (Figure from Ref. [28]); b) Noise in charge-ordered phase of Nd$_{0.5}$Ca$_{0.5}$MnO$_3$ as a function of bias. The corresponding $I-V$ curve is also shown. (Figure from Ref. [29]).

Transport in a new system of metallic dot arrays[32] exhibit the same threshold phenomena due to Coulomb blockade. Although no noise measurement is known in this system, numerical simulations[33] have brought out many interesting features of noise in collective transport. The current near the threshold is given by $I \sim (V - V_T)^\zeta$ where $\zeta$ is an exponent. At larger bias $I - V$ becomes linear. It was found that the crossover to ohmic behaviour in 2D is accompanied by a change in the flow from 2D meandering to straight 1D flow. Interestinly, at low bias noise shows a normal 1/f-type feature but in ohmic regime, it shows a characteristic narrow-band type peak. Such signatures would be easily observable in experimental noise spectra.

*Systems without threshold.* Noise peaks are also found in field-induced phase transitions which are accompanied by large nonlinearity in transport (see Fig. 7b). Recent examples include melting of charge-ordered states in manganites[29], crossover from ohmic regime to SCLC regime in organic semiconductors[34]. It is seen in Fig. 7b that unlike in the above, the noise starts increasing much earlier than the onset of nonlinearity in $I - V$ curve indicating that $b > 1$.

## 5. DISCUSSION AND OPEN ISSUES

Much of the phenomenalogical description of noise in non-ohmic regimes given in sections above lack theoretical support. For example, it will be nice to have some basis for Eqns. (6) and (7) or, for the fact that $\gamma \leq 0$ in many disordered systems. Apart from these, there are several fundamental problems associated with the noise in what is basically field-driven nonequilibrium regimes. One of the earliest theoretical discussion of fluctuation in nonlinear systems was by Van Kampen[35]. Some conceptual issues are as follow:

*i) Fluctuation-dissipation theorem:* The theorem is known to be valid only for equilibrium systems. However, in last decade there have been attempts to extend it in nonequilibrium conditions. See Ref. [36] and references therein. According to these, asymmetry of fluctuations of energies can be related to the dissipation energies required to maintain a nonequilibrium steady state. *ii) Fluctuation distribution:* Many work[20, 27, 37] have pointed out that fluctuations tend to be nongaussian particularly in media with large disorder. *iii) Coherence length:* The coherence length of fluctuations is normally assumed to be of the order of microscopic lengths. But, in fluids in in nonequilibrium states[38] and in CDW states[28] the length could be of macroscopic order. The same was invoked to explain the huge increase in noise in the Joule regime of composites[15].

## ACKNOWLEDGMENTS

We thankfully acknowledge assistance of Arindam Chakrabarti in sample preparation and data analysis. We also thank our collaborators S. De and P. Nandy, and A. Carbone for sending us a preprint in advance.